\documentclass[prd,twocolumn,amsmath,amssymb,floatfix,superscriptaddress]{revtex4}

\usepackage{graphicx}
\usepackage{amssymb}
\usepackage{amsmath}
\usepackage{color}
\usepackage{ wasysym }
\usepackage{hyperref}
 \usepackage{tabu}

\def\barray{\begin{array}}
\def\earray{\end{array}}
\def\be{\begin{equation}}
\def\ee{\end{equation}}
\def\ben{\begin{equation} \nonumber}
\def\een{\end{equation}}
\def\ban{\begin{eqnarray*}}
\def\ean{\end{eqnarray*}}
\def\ba{\begin{eqnarray}}
\def\ea{\end{eqnarray}}

\def\({\left(}
\def\){\right)}

\def\apj{ApJ}

\def\mnras{MNRAS}

\def\aj{AJ}

\def\physrep{Phys. Rep.}

\def\apjs{ApJS}

\graphicspath{{./fig/}}

\begin{document}

\title{New parameterization for unified dark matter and dark energy }
\author{Zahra Davari}
\email{z.davaridolatabadi@sci.basu.ac.ir }
\affiliation{Department of Physics, Bu-Ali Sina University, Hamedan
	65178, 016016, Iran}

\author{Mohammad Malekjani}
\email{malekjani@basu.ac.ir}
\affiliation{Department of Physics, Bu-Ali Sina University, Hamedan
	65178, 016016, Iran}

\author{Michal Artymowski}
\email{michal.artymowski@uj.edu.pl}
\affiliation{Institute of Theoretical Physics, Faculty of Physics, University of Warsaw, ul. Pasteura 5, 02-093 Warsaw, Poland }
\affiliation{Institute of Physics, Jagiellonian University, {\L}ojasiewicza 11, 30-348 Krak{\'o}w, Poland }

\begin{abstract}
In this paper we investigate a new phenomenological parameterization for unified dark matter and dark energy based on the polynomial expansion of the barotropic equation of state parameter $w$. Our parameterization provides well-behaving evolution of $w$ for both small and big redshifts as well as in the far future. The dark fluid described by our parameterization behaves for big redshifts like a dark matter. Therefore one can parameterize dark energy and dark matter using a single dark fluid, like in the case of the Chaplygin gas. Within this parameterization we consider 2 models: one with DE barotropic parameter fixed to be $-1$ and the second one, where $w \neq -1$ is chosen to match the best fit to the data. We study main cosmological properties of these models at the expansion and perturbation levels. Based on Markov chain Monte Carlo method with currently available cosmic observational data sets, we constrain these models to determine the cosmological parameters at the level of background and clustering of matter. We consider the interaction between DM and DE which directly affects the evolution of matter and its clustering. Our model appears to be perfectly consistent with the $\Lambda$CDM model, while providing unification of DE and DM. 
\end{abstract}
\maketitle

\section{Introduction} \label{sec:intro}
Various evidences from independent cosmic observations such as measurements of the rotation curves of spiral galaxies\citep{Persic:1995ru, Borriello:2000rv}, dynamics of galaxy clusters\citep{Frenk:1995fa} and cosmic structure formation \citep{Primack:1996dh}  show that there is roughly six times more cold dark matter (CDM) than can be afforded by the baryonic matter in the cosmic matter budget making up of order  $\sim 30\%$ of critical density\citep{Planck2015_XIII}. In addition to this clustering dark component, the astronomical observations including e.g. supernovae Type Ia (SNIa) \citep{Riess1998,Perlmutter1999,Kowalski2008}, cosmic microwave background (CMB) fluctuations \citep{Komatsu2009,Komatsu2011,Jarosik:2010iu,Planck2015_XIII}, large-scale structure (LSS) by Sloan Digital Sky Survey (SDSS) \citep{Tegmark:2003ud}, baryonic acoustic oscillation (BAO) \citep{Percival2010,Cole:2005sx,Eisenstein:2005su,Reid:2012sw,Blake:2011rj} and galaxy clustering provide evidences for the existence of the so-called dark energy (DE) - an exotic fluid with sufficiently negative pressure, which causes the late-time accelerated expansion of the Universe. 

Despite a lot of studies on DM and DE, their physical properties,origin and nature are yet unknown. In literature, many candidates have been suggested for dark matter such as axions \citep{Shellard:1998mi}, the lightest supersymmetric particle (LSP) like neutralinos \citep{Falvard:2002ny} and the Kaluza–Klein particles \citep{Bottino:2005xy} that are Weakly Interacting Massive Particles (WIMPs) \citep{Appelquist:2000nn,Randall:1999ee}. Also for DE, the most natural candidate is  cosmological constant with constant equation of state (EoS) $w_{\Lambda}=-1$, but there is a discrepancy of some 120 orders of magnitude between its theoretical and observed values known as the fine-tuning problem \citep{Weinberg:1988cp,Carroll:2000fy, Padmanabhan2003, Copeland2006}. For this reason, other candidates such as quintessence and k-essence models with time varying EoS parameter have been suggested. Quintessence models involve canonical kinetic terms of the self-interacting scalar field \citep{Caldwell:1997ii} and k-essence models contain exotic scalar fields with non-canonical (non-linear) kinetic terms which typically lead to a negative pressure. Generally speaking, it has been proposed that we cannot entirely understand the nature of DE before the establishment of a complete theory of quantum gravity \citep{Witten:2000zk}. One can ask if it's possible to obtain a simple model, in which a single dark fluid (DF) behaves as both dark matter and dark energy \citep{Silva:2003bs}. This attractive dark fluid with barotropic equation of state (EoS) $w$ (which is the ratio of pressure to energy density) can unify DM and DE and explain both the accelerated and decelerated expansions at late and early times respectively. In other words the barotropic EoS parameter of DF acts like DM EoS parameter ($w_{\rm m}\sim 0$) at high redshifts and behaves like DE EoS parameter ($w_{\rm de}<-1/3$) at low redshifts. This dual role of DF is the most interesting and surprising property in these scenarios. For this fluid, the coincidence problem of $\Lambda CDM$ (i.e., why we live in a particular era during which both dark components 
are of the same order of magnitude at the present whereas they were so different in most of the past evolution of the Universe) is resolved\citep{Sahni:2000zv}. One particular case of DF which unifies DM and DE is the so called generalized Chaplygin gas (gCg), introduced by Kamenshchik \citep{Kamenshchik:2001cp} and then developed in \citep{Bento:2002ps}.\\

On the other hand, one possible way to study the EoS parameter of DE models is via  parameterizations. In literature, we can find different EoS parameterizations for DE in which the EoS parameter of DE is defined as a function of cosmic redshift ($w(z)$). The simplest and earliest EoS parameterizations are introduced based on the Taylor expansion of EoS parameter of DE, $w_{de}$,  with respect to redshift $z$  as: $w_{de}(z)=w_{0}+w_{1}z$ \citep{Maor:2000jy,Riess:2004nr} and with respect to $(1-a)$ as: $w_{de}(z)=w_{0}+w_{1}\frac{z}{1+z}$ where $a = 1+z$ is a scale factor of the FRW metric. The second parameterization is a well-known Chevallier-Polarski-Linder (CPL) parameterization proposed by  \cite{Chevallier2001, Linder2003}. In addition, some purely phenomenological parameterizations have been introduced in recent years. For instance $w_{\rm de}(z)=w_0+w_1z/(1+z)^{\gamma}$, where $\gamma$ fixes to 2 \citep{Jassal:2004ej}. Moreover, the power law $w_{\rm de}(a)=w_0+w_1(1-a^{\alpha})/\alpha$ and logarithmic $w_{\rm de}(a)=w_0+w_1 \ln{a}$  parameterizations have been suggested  \citep{Barboza:2009ks,Efstathiou:1999tm}. Another phenomenological parameterization is the Wetterich parameterization $w_{\rm de}(z)=w_0/[1+b\ln{(1+z)}]^{\alpha}$, where $\alpha$ is fixed to  $1$ or $2$ \citep{Wetterich:2004pv}. The important note is that although the CPL formula is a well-behaved parameterization at early ($a\ll 1$) and present ($a\sim 1$) epochs, it diverges when the scale factor goes to infinity at far future. All of the above parameterizations are introduced to describe the evolution of EoS parameter of DE. Based of our knowledge, there is still no specific parameterization to describe the barotropic EoS parameter of DF consists of  DM and DE. In this work we introduce a new parameterization for barotropic EoS parameter of DF ( hereafter, \textit{DF parameterization}). Using this parameterization (see section (\ref{sec: 2}) for complete description), the barotropic EoS parameter of DF can tend to $w=0$ at early matter dominated Universe  and $w<-1/3$ at late time accelerated Universe.
This paper is organized as follows. In Sec.	 (\ref{sec: 2}) we introduced the DF  parameterization and study the evolution of its EoS parameter. In Sect. (\ref{sec:3}) the cosmological background evolution based on DF parameterization is investigated. We study the redshift evolution of main cosmological quantities using DF parameterization. Using the latest cosmological data in background level including data from joint light-curve analysis (JLA) supernovae, CMB, BAO, big bang nucleosynthesis  (BBN) and Hubble expansion rate, in the context of Markov Chain Monte Carlo (MCMC) algorithm we perform a joint likelihood statistical analysis in order to constraint the free parameters of DF parameterization. In Sect. (\ref{sec:5}), we investigate the growth of matter perturbations using DF parameterization and then perform a likelihood analysis using growth rate of perturbations to place constrain on the parameters of model in perturbation level and obtain their best fit values. Finally, we summarize our results in Section  (\ref{sec:6}).

\begin{figure}
	\begin{center}
		\includegraphics[height=5cm]{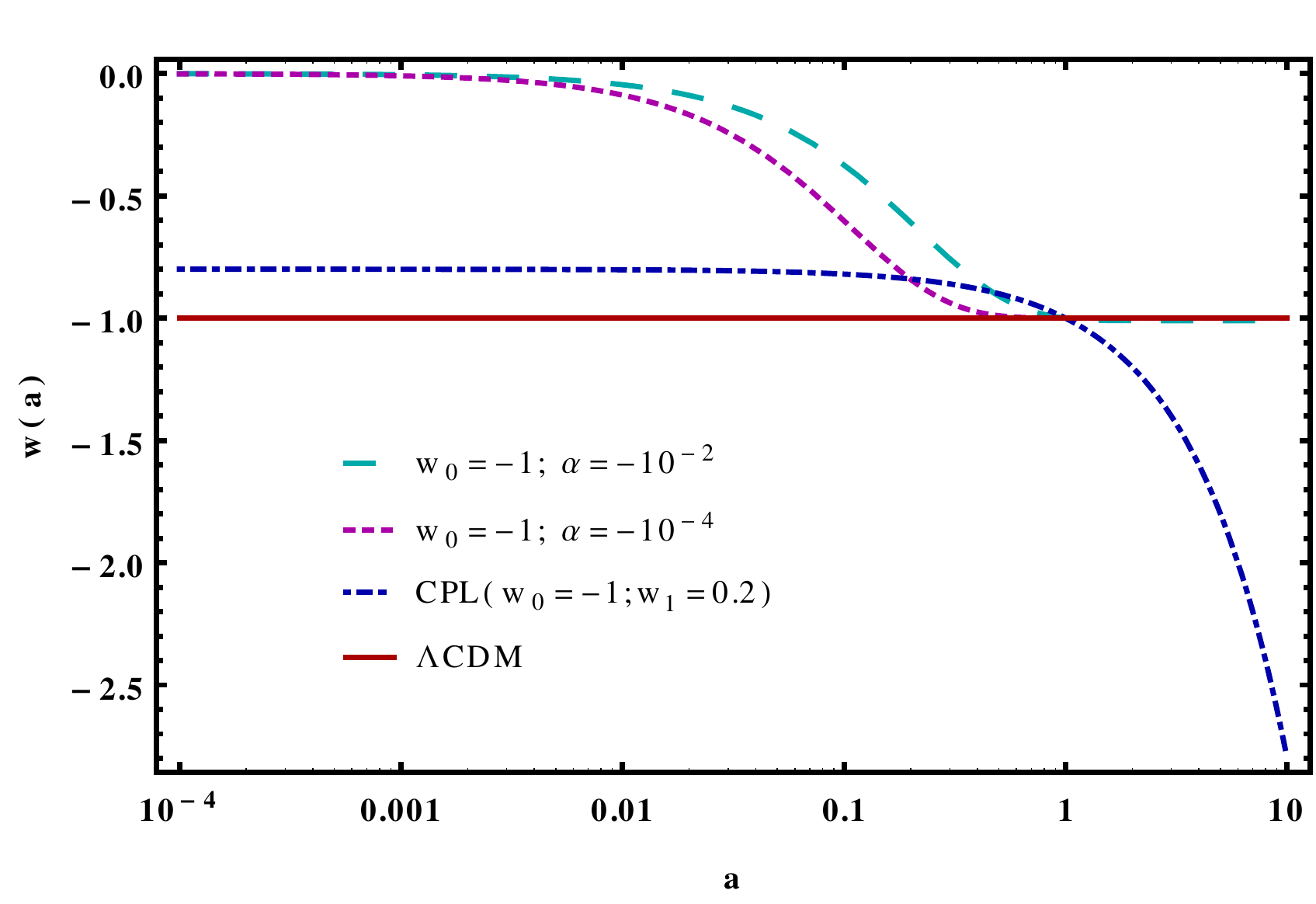}
		\caption{The evolution of the EoS parameter of unified DF and CPL parameterization as a function of scale factor $a$. The EoS parameter of CPL parameterization goes to infinity at far future while the EoS parameter of unified DF coincides the $\Lambda$CDM one at today and future times.}
	\end{center}\label{fig:1}
\end{figure}
\section{DF parameterization} \label{sec: 2}
First, let us consider a simple parameterization for the barotropic EoS parameter of DF as 
\begin{equation}
w(a) =  w_{0} + \sum_{n=1}^{N} w_{n} (1-a)^{n} \, , \label{eq:wfinite}
\end{equation}
where $a$ is scale factor normalized to $1$ at present time,  $w_{n}$ are constant coefficients and $w_{0}$ is a present-day barotropic parameter. The issue is that this parameterization leads to divergences  of $w(a)$ in the far future, i.e. in the $a \to \infty$ limit. In principle this is not an issue that would experimentally exclude the model. Quite obviously one can only measure past values of $w$, so its future behaviour cannot be a subject of experimental verification. Nevertheless a parameterization with non-stable $w$ could be considerd as decoupled from the predictions of most of the theoretical models of DE. Starting from DE models motivated by the field theory, like quintessence, to $f(R)$ and scalar-tensor theories, one usually obtains rather constant value of the EoS parameter in the far future. Thus, a part of our motivation is to include in our analysis a connection between a phenomenological parameterization and more fundamental theories of DE. In order to solve the problem of unstable $w$ one can secure the existence of finite  and slowly evolving $w(a)$ for arbitrary big, but finite value of $a$ by assuming that $w(a)$ has a stationary point at some $a = a_{s}$. The maximal order of a stationary point in the case of Eq. (\ref{eq:wfinite}) is equal to $N-1$, which leads to $N-1$ constrains on $w(a)$, namely
\begin{equation}
\frac{dw}{da} = \frac{d^2w}{da^2} = \ldots = \frac{d^{N-1}w}{da^{N-1}} = 0 \, . \label{eq:conditions}
\end{equation}
We have obtained $N-1$ independent equations that constrain $N+2$ parameters (including $N+1$ different $w_n$ and $a_s$), which gives $3$ independent parameters of such a model. Note that this number is $N$-independent. From Eq. (\ref{eq:wfinite}) and (\ref{eq:conditions}) one can find
\begin{equation}
a_{s} =\left(N\frac{w_{N}}{w_1}\right)^{\frac{-1}{N-1}}\, , 
\quad w_{n} = (-1)^{n+1}\frac{(N-1)!}{n!(N-n)!}\left(N  \frac{w_{N}}{w_1}\right)^{\frac{n-1}{N-1}} \, . \label{eq:wn}
\end{equation}
For an arbitrary $w_{ N}$ the barotropic EoS parameter $w(a)$ does not necessary converges in the $N \to \infty$ limit. Note however that for
\begin{equation}
w_{ N} =  \sigma (w_1/N)^{N} \, ,
\end{equation}
where $\sigma$ is any positive constant, one can obtain a $N \to \infty$ limit of the theory, which gives
\begin{equation}
w_{ N} \to 0 \, ,\qquad a_{s} \propto \frac{N}{w_1} \to \infty \, , \qquad w(a) = w_{0} + 1 - e^{-w_{1} (a-1)} \, . \label{eq:wflat} 
\end{equation}
This leaves us with two parameters of the theory. With growing $N$ one obtains bigger $a_{s}$, which secures lack of divergence of $w(a)$ for bigger values of $a$. The $N \to \infty$ limit guarantees lack of divergences for all $a$. It is easy to show that any barotropic EoS parameter $w(a)$ of the form
\begin{equation}
w(a) = w_{0} + f(1 - e^{-w_{1} (a-1)}) \, , \label{eq:wgeneral}
\end{equation}
where $f(a)$ is continuous, differentiable, and well defined function for all $a$, satisfies conditions (\ref{eq:conditions}) in the $N \to \infty$ limit. We have normalized the scale factor to be equal to 1 today and we have assumed $f(a=1)=0$ in order to obtain $w_0$ as a present value of a barotropic parameter of DF. Such a theory may have arbitrary number of free parameters within the $f(x)$ function. Again, let us emphasize that the infinite order stationary point in $a \to \infty$ secures a finite value of $w(a)$ for all values of the scale factor. Therefore one avoids divergences of $w(a)$, which are otherwise present for this type of parameterization. Based on Eq. (\ref {eq:wgeneral}), let us assume one of the simplest forms of $f(a)$, namely
\begin{equation}
w(a) = w_{0} +\alpha\left( 1 - e^{-w_{1} (a-1)}\right) \label{eq:wflat1} \, ,
\end{equation}
where $\alpha$ is a constant. According to Eq. (\ref{eq:wflat1}), one can obtain the following three limits for the evolution of the barotropic parameter $w(a)$:
\[w(a) = \left\{
\begin{array}{l l l}
w_{0}+\alpha(1-e^{w_{1}}) \quad \quad a \to 0 \\
w_{0} \quad\quad \quad\quad\quad\quad\quad\quad a \to 1\\
w_{0}+\alpha \quad \quad\quad\quad\quad\quad a \to \infty\\
\end{array} \right. \]
where $a \to 0$ ($a \to \infty$) represents far past (future) of the Universe. Note that both, $a\to 0$ and $a \to \infty$ limits lead to a constant and finite values of $w(a)$, which are significantly different from each other. This leads to the rapid transition period between phases of quasi-constant values of $w$, which has been investigated in the context of different parameterizations of inflation and DE in \cite{Artymowski:2016ikw, Carrasco:2015rva, Dimopoulos:2017zvq, Ford:1986sy, Kunimitsu:2012xx}. In principle we want to obtain a smooth transition between some initial, almost constant value of the barotropic parameter denoted as $w_{\rm in}$ and a final one, which is $w_0+\alpha$. In order to obtain such an evolution of $w$ let us note that for 
\begin{equation}
w_1 = \log\left(\frac{w_0 + \alpha - w_{\rm in}}{\alpha}\right) \label{eq:w1flat}
\end{equation}
one finds $w \to w_{\rm in}$ for $a\to 0$.  In such a case from Eq. (\ref{eq:wflat1}) and Eq. (\ref{eq:w1flat}) one finds the following form of the barotropic parameter
\begin{equation}
w = w_0 + \left(1 - \left(\frac{w_0+\alpha-w_{\rm in}}{\alpha}\right)^{1-a} \right) \, . \label{eq:w41}
\end{equation}
In this paper we investigate the case of $w_{\rm in} = 0$. Nevertheless one could use this parameterization to describe the transition between a massless scalar field and DE (for $w_{\rm in}=1$) \cite{Dimopoulos:2017zvq} or between radiation-like evolution of a fluid and DE ($w_{\rm in} = 1/3$)
\footnote{The case of the transition between $w=1/3$ and $w=-1$ can be realized for the K-essence \cite{Garcia:2012kr}}. Note that one needs $\alpha<0$ in order to obtain those limits. 

Based on Eq. (\ref{eq:w41}), in Fig. (\ref{fig:1}) we show the evolution of the EoS parameter of DF $w(a)$ as a function of scale
 factor for $w_0 = -1$ and two different values of free parameter $\alpha$. Note that for our parameterization of $w(a)$ one can always find $a$, for which $w$ vanishes, meaning that the barotropic parameter of the DF behaves as pressure-less DM. On the other hand, at present and future times, $w(a)$ tends to $-1$ which means that DF acts like cosmological constant. For comparison, we also show the EoS parameter of CPL parameterization and the constant EoS parameter of $\Lambda$CDM model. We see that the CPL parameterization diverges when the scale factor goes to infinity at far future.

\section{ DF parameterization against geometrical observations}  \label{sec:3}

In this section, we obtain the basic equations governing the evolution of background cosmology within DF parameterizations. Then using the observational data in the background level, we perform the statistical MCMC analysis to put constraints on the cosmological parameters in the context of DF parameterization. Finally we show the evolution of main cosmological quantities describing the evolution of background cosmology in DF parameterization.
\subsection{Basic equations }\label{basic_eq}
In this section we use Eq. (\ref{eq:w41}) with $w_i = 0$ as a parameterization describing the barotropic EoS parameter of DF (denoted in here as DF parameterization) to study the evolution of Hubble flow in the spatially flat Friedman–Robertson–Walker  (FRW) Universe. We assume that the Universe is filled with baryonic matter, radiation and unified DF. Then, the first Friedmann equation takes the following form
\begin{equation}
H^{2}=\frac{1}{3m_{p}^{2}}( \rho_{ \rm b}+ \rho_{\rm r}+ \rho_{\rm df})\;, \label{frid}
\end{equation}
where $m_p^2=\frac{1}{8\pi G}$ is the reduced Planck mass and  $\rho_{\rm b}$, $\rho_{\rm r}$ and  $\rho_{\rm df}$ are the energy densities of baryons, radiation and DF respectively. In the absence of interactions among the above fluids the evolution of the energy density  as a function of the cosmic scale factor $a(t)$ is characterized by the continuity equation as follows
\begin{equation}
\frac{d \rho_{i}}{d a} + \frac{3}{a}(1+w_{i}(a))\rho_{i} = 0 \, , \label{eq:cont}
\end{equation}
where $\rho_{\rm i}(a)$ is the energy density of radiation ($w_{\rm i}=1/3$), baryonic matter ($w_{\rm i}=0$) and DF is given via Eq. (\ref{eq:w41}) respectively. It is easy to  derive the evolution of the energy density of baryons and radiation as $\rho_{\rm b}=\rho_{b0} \, a^{-3}$ and $\rho_{\rm r}=\rho_{r0} \, a^{-4}$ respectively. Also inserting Eq. (\ref{eq:w41}) in Eq. (\ref{eq:cont}), we can obtain the evolution of $\rho_{\rm df}$  as
\begin{equation}
\rho_{\rm df}(a)=\rho_{df0} \ e^{-3\int_{1}^{a} \frac{1+w(x)}{x} dx}  \;.  \label{eq:roo}
\end{equation}
Using the definition of the dimensionless energy density parameter $\Omega_{i}=\frac{\rho_{i}}{\rho_{cr}}$, where $ \rho_{cr}=3m_{p}^{2}H^2$ is the critical  energy density,  the dimensionless  Hubble parameter,$E=\frac{H}{H_{0}}$ takes the following form 
\begin{equation}
E=\sqrt{\Omega_{b0}a^{-3}+\Omega_{r0}a^{-4}+\Omega_{df0} \ e^{-3\int_{1}^{a} \frac{1+w(x)}{x} dx} }\;.  \label{eq:roo2}
\end{equation}
Applying Friedman equation, which takes the form of $\sum{\Omega_{i}}=1$, we can write $\Omega_{df0}=1-\Omega_{b0}-\Omega_{r0}$, where $\Omega_{\rm r0}=2.469\times 10^{-5}h^{-2}(1.6903)$ is the energy density of radiation (photons+relativistic neutrinos) and $h=H_0/100$ \citep{Hinshaw:2012aka}.
 Notice that in the case of our model DF is considered as the unification of DM and DE and therefore the DM energy density is not explicitly included in the Hubble flow. In order to obtain the evolution of DM and DE separately, one needs to decompose the energy density of DF as 
\begin{equation}
\rho_{\rm df}=\rho_{dm}+\rho_{de}\;.  \label{eq:roo1}
\end{equation}
According to energy-momentum conservation equation one obtains the continuity equation for DF as
\begin{equation}
\dot{\rho}_{\rm df}+3H(1+w)\rho_{\rm df}=0 \;.
\end{equation}
Then, the continuity equations for DM and DE are respectively given by
\begin{eqnarray}
&&\dot{\rho}_{dm}+3H\rho_{dm}=Q \;,\\
&&\dot{\rho}_{de}+3H(1+w_{de})\rho_{de}=-Q\;,
\end{eqnarray}
where Q is the interaction parameter between DM and DE (\cite{Amendola:1999er,delCampo:2006vv,Poitras:2013zja}). Note, that unified models of DM and DE have certain issue related to the behavior of the speed of sound, namely they may lead to the production of unphysical oscillations \citep{Sandvik:2002jz}. The unified models behave like DM at early times and therefore their sound speed is vanishing. As one approaches the present time, the unified models behave like DE with negative pressure resulting a large sound speed which produces oscillations or blow up in the power spectrum \citep{Sandvik:2002jz}. In unified models this is unavoidable feature unless we identify the DM and DE components of one fluid. In our case it is natural to assume, that DM and DE are interacting as they are both considered as a single fluid. Due to the interaction one obtains a dissipation of energy between DM and DE, which can be estimated phenomenologically as \citep{Marcondes:2016reb,Yang:2013hra}
\begin{eqnarray}\label{eq:Q}
Q=3aH(\xi_1\rho_{dm}+\xi_2\rho_{de})\;,
\end{eqnarray}
where $\xi_1$ and $\xi_2$ are the coupling coefficients and they can be determined by observations. The energy flow from DE to DM is defined by $Q>0$, and oppositely $Q<0$ shows the energy flow from DM to DE.
\\* 

\begin{figure}
	\begin{center}
		\includegraphics[height=6cm,width=7.5cm]{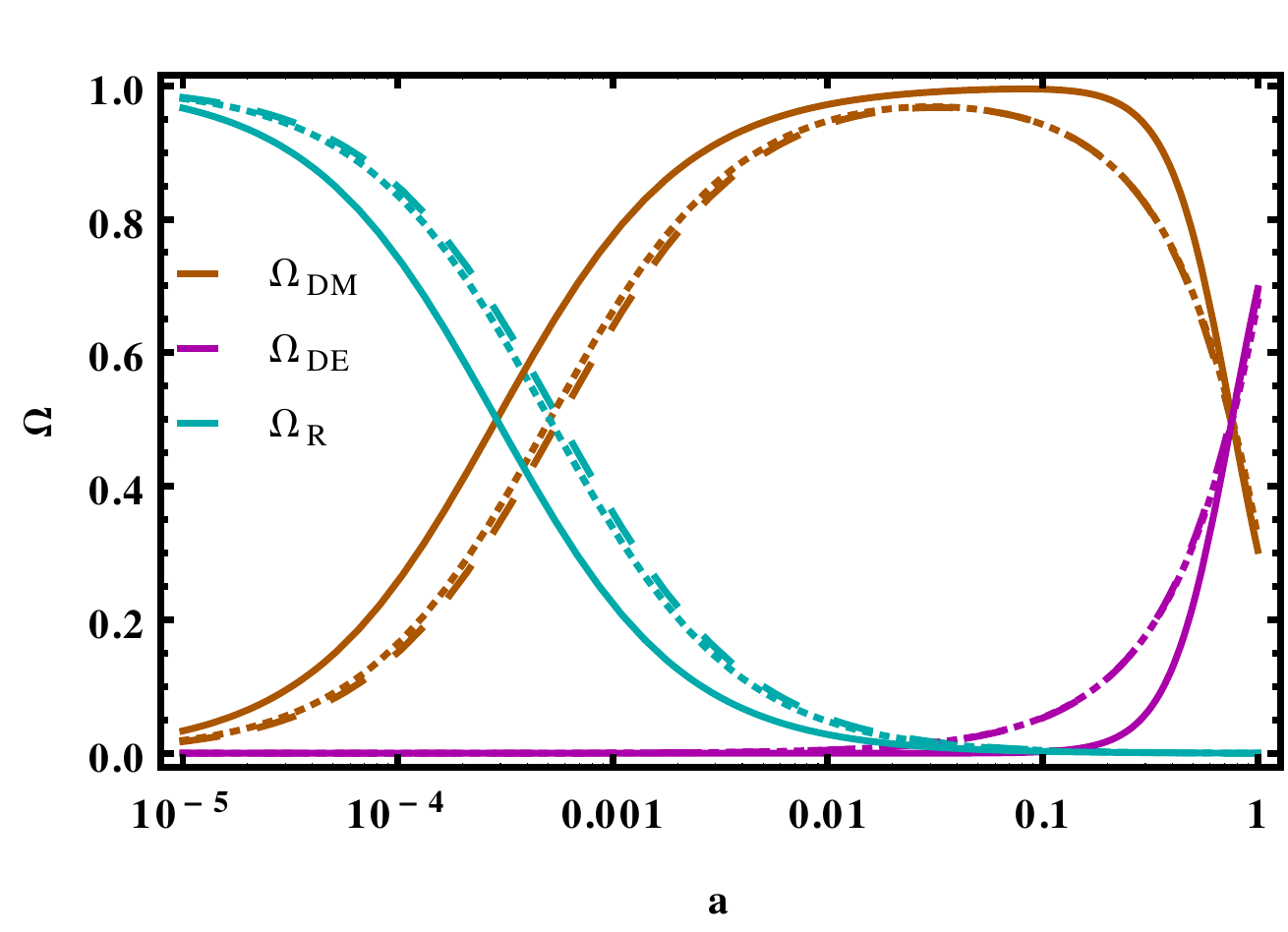}
		\caption{ Evolution of the fractional energy density of radiation (green curves), non-relativistic matter (brown curves) and DE component (pink curves) in terms of cosmic scale factor for different DF parameterizations and standard $\Lambda$CDM cosmology. Dashed and dotted curves stand for case 1  and case 2 of DF parameterization respectively. The solid curves represent the concordance $\Lambda$CDM model. In all models, we use the best fit values from Tab. (\ref{tab:res_all}).}\label{fig: 2}
	\end{center}
\end{figure}

Using Eq. (\ref{eq:roo1}) and barotropic equation for DF $P_{\rm df}=w(a)\rho_{\rm df}$, assuming $P_{\rm df}=P_{de}$ ( since $P_{dm}=0$), we can obtain $\rho_{dm}=(\frac{w_{de}(a)-w(a)}{w_{de}(a)} )\rho_{\rm df}$ and $\rho_{de}=\frac{w(a)}{w_{de}(a)}\rho_{\rm df}$. Hence the dimensionless energy densities of DM and DE can be obtain as:  $\Omega_{dm}= \frac{w_{de}(a)-w(a)}{w_{de}(a)} \Omega_{\rm df}$ and $\Omega_{de}=\frac{w(a)}{w_{de}(a)}\Omega_{\rm df}$. The energy density of DM and DE can be obtained as follows
\begin{equation}
\Omega_{dm}=\frac{\Omega_{df0}(w_{de}-w(a))e^{-3\int_{1}^{a} \frac{1+w(x)}{x} dx}}{w_{de}E^{2}(a)}\;,
\end{equation}
\begin{equation}
\Omega_{de}=\frac{\Omega_{df0}w(a)e^{-3\int_{1}^{a} \frac{1+w(x)}{x} dx}}{w_{de}E^{2}(a)}\;.
\end{equation}
In the rest of paper we consider two different cases for DE. First we assume DE as cosmological constant with $w_{de}=-1$ (case 1). Second, we consider DE as quintessence model with constant $w_{de}$ differs from $-1$ (case 2) in order to provide the best fit to the data (see Tab. (\ref{tab:res_all}) for details). Let us note that both cases are considered within the parameterization of the DF presented in the Eq. (\ref{eq:w41}).

\subsection{Geometrical observations and cosmological constraints} 

Now using the background expansion data including those of {JLA supernovae binned sample \citep{Betoule:2014frx,Escamilla-Rivera:2016aca}}, BAO \citep{Beutler:2012px, Xu:2012fw, Anderson:2013zyy, Blake:2011rj}, {Planck data for the position of CMB acoustic peak \citep{Shafer:2013pxa}}, BBN \citep{Serra:2009yp}, Hubble data \cite{Moresco:2012jh, Gaztanaga:2008xz, Blake:2011rj, Anderson:2013zyy}, we implement a statistical MCMC analysis for two classes of DF parameterization described in Sect. (\ref{basic_eq}). For more details regarding the MCMC method used in this work, we refer the reader to \cite{Mehrabi:2015hva} \citep[see also][]{Basilakos:2009wi,Hinshaw:2012aka,Mehrabi:2015kta,Mehrabi:2016exz,Malekjani:2016edh}. In this section we have used the following sets of data: {31 distinct points for JLA binned sample data}, 37 points for Hubble data \citep[see Table 3 of][]{Sola:2016jky}. The BAO data include 6 distinct measurements of the baryon acoustic scale \citep[see Tab.1 of][]{Mehrabi:2015hva}. We use the Plank data for the position of CMB acoustic peak in \cite{Shafer:2013pxa}. The Big Bang Nucleosynthesis (BBN) provides  data point which constrains $\Omega_{\rm b}^{(0)}$ \citep{Serra:2009yp}. The total likelihood function is the product of the  individual likelihoods for each experiments as
\begin{equation}\label{eq:like-tot}
{\cal L}_{\rm tot}({\bf p})={\cal L}_{\rm sn} \times {\cal L}_{\rm bao} \times {\cal L}_{\rm cmb} \times {\cal L}_{h} \times
{\cal L}_{\rm \rm bbn}\;,
\end{equation}
so the total chi-square ($\chi^2_{\rm tot}$) is given by sum of individual chi-squares:
\begin{equation}\label{eq:like-tot_chi}
\chi^2_{\rm tot}({\bf p})=\chi^2_{\rm sn}+\chi^2_{\rm bao}+\chi^2_{\rm cmb}+\chi^2_{h}+\chi^2_{\rm bbn}\, ,
\end{equation}
where the statistical vector $p$ contains free parameters of cosmological model. This vector for $\Lambda$CDM cosmology contains $\{\Omega_{b0}$,$\Omega_{dm0}$,$H_{0}\}$. For DF parameterization (case 1), the vector $p$ contains $\{\Omega_{b0}$,$H_{0}$,$w_{0}$,$\alpha$\} and for DF parameterization (case 2) the vector $p$ includes $\{\Omega_{b0}$,$H_{0}$,$w_{0}$,$\alpha$,$w_{\rm de}\}$. Notice that in the case 2, we have one more free parameter ($w_{\rm de}$) than case 1, since in the case 2 we consider DE as a quintessence with unknown EoS parameter $w_{\rm de}$,  while in case 1, DE is considered as cosmological constant $\Lambda$ with constant EoS $w_{\rm \Lambda}=-1$. In this analysis we fix the energy density of radiation ( photons+relativistic neutrinos) as $\Omega_{\rm r0}=2.469\times 10^{-5}h^{-2}(1.6903)$ where $h=H_0/100$ \citep{Hinshaw:2012aka}.

\begin{figure*}
	\includegraphics[width=8cm]{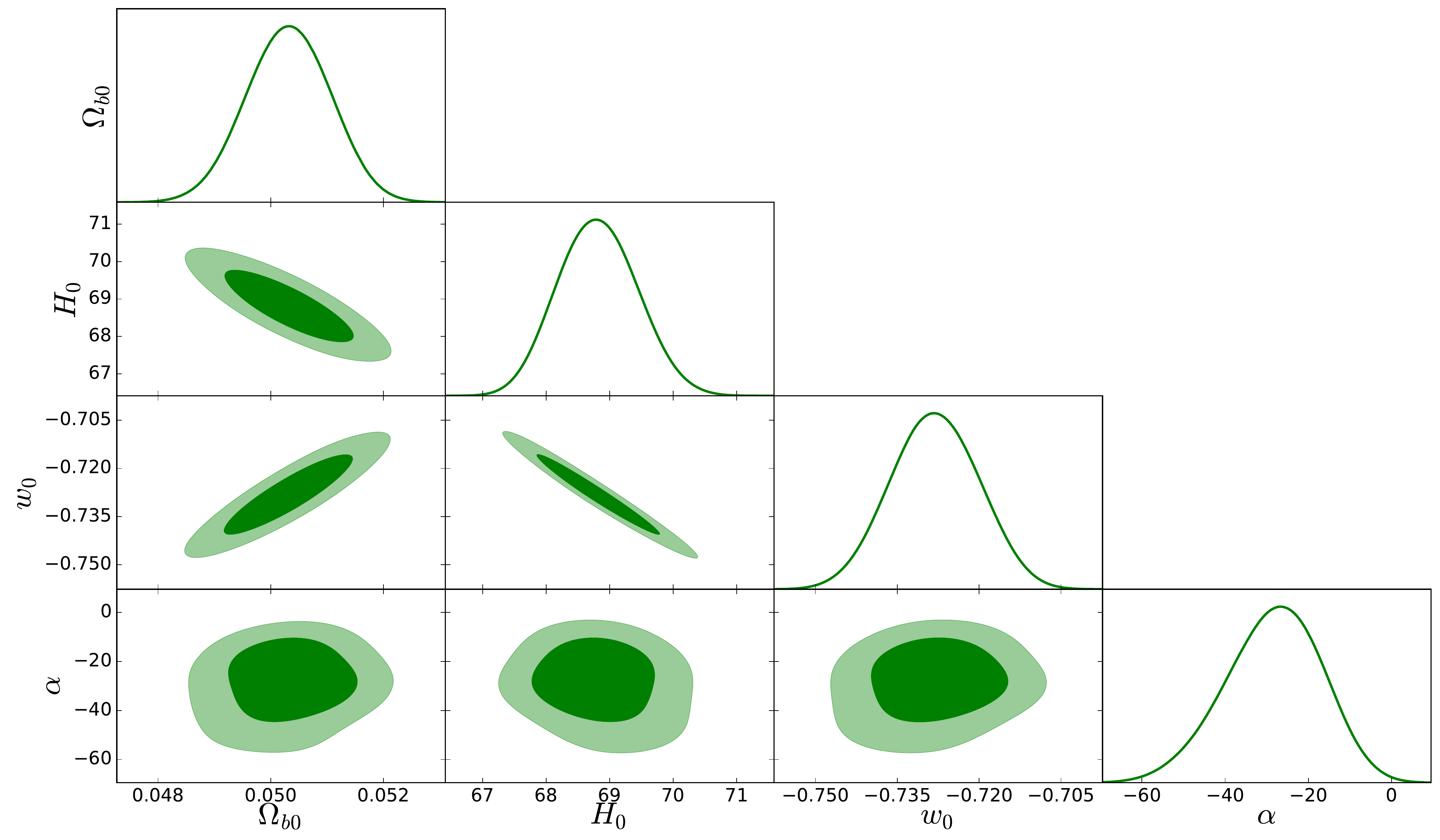}
	\includegraphics[width=8cm]{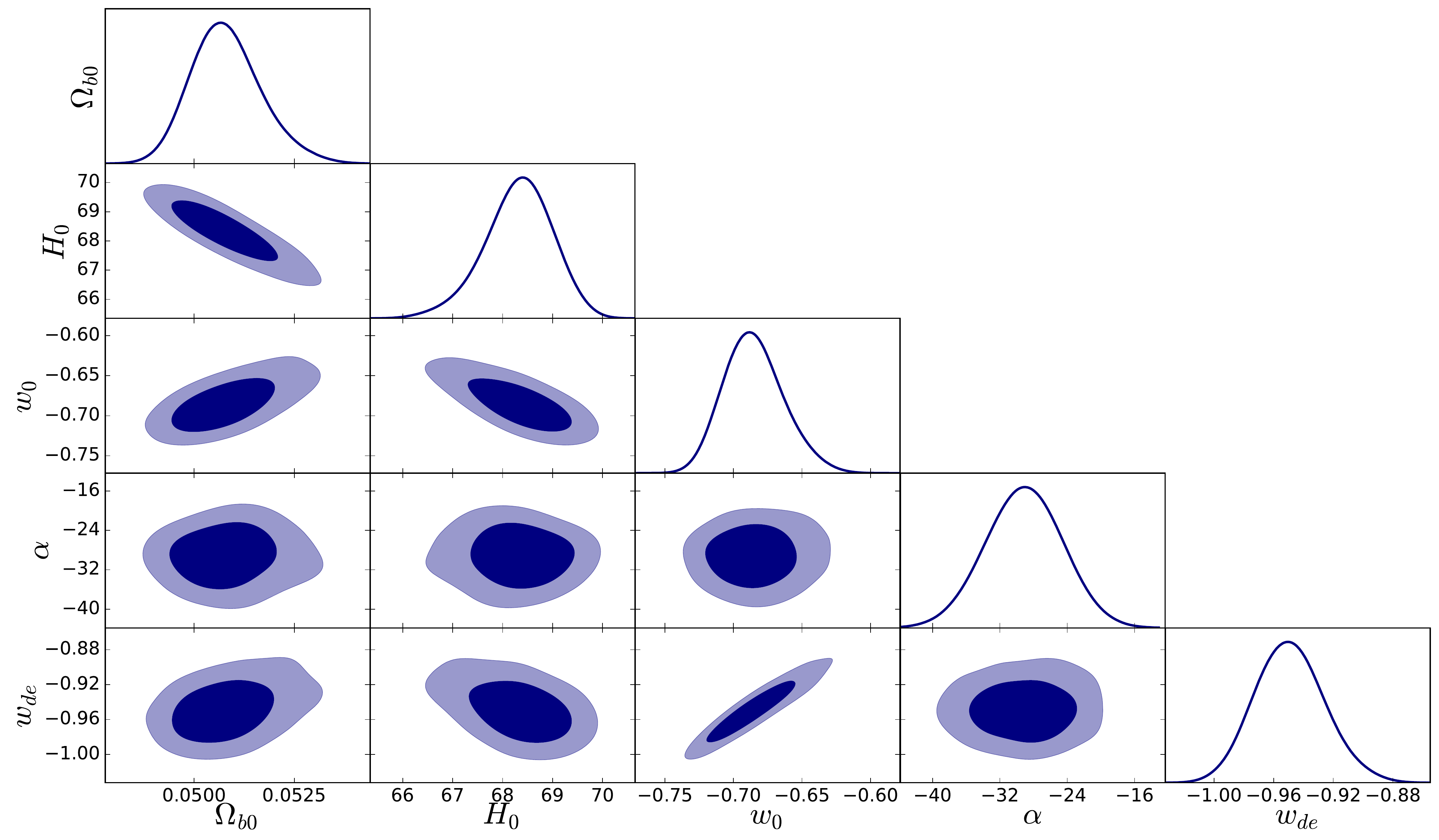}
	\includegraphics[width=8cm]{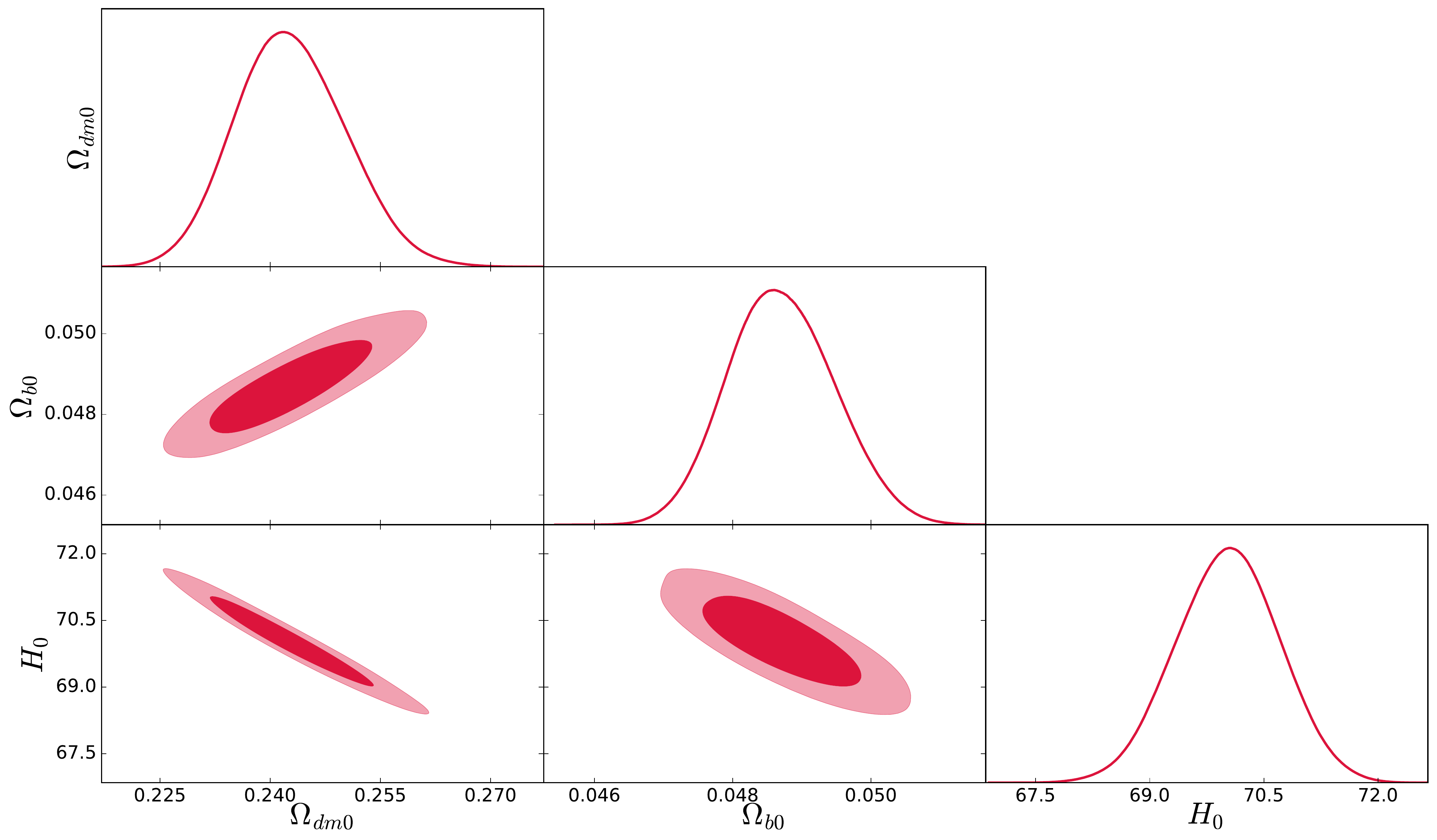}
	\caption{ The $1\sigma$ and $2\sigma$ confidence levels and maximum likelihood function for various cosmological {parameters} planes. Upper left (upper right) panel stands for case 1 (case 2) and bottom panel stands for $\Lambda$CDM cosmology.}
	\label{figc2}
\end{figure*}

In chi-square analysis it is clear to conclude that a model with lower value of $\chi^2_{\rm min}$ is better fitted to observational data compare to other models. However, this result is valid if the number of free parameters of models are equal. In an other word, this analysis is no longer valid for comparing different models with different number of free parameters. Hence we use another statistical tests the so called Akaike information criteria (AIC)  and BIC to compare DF parameterizations with observations. Notice that in $\Lambda$CDM cosmology we have 3 free parameters, in the case of  DF parameterization (case 1) there are 4 free parameters and in case (2) we have 5 free parameters. The AIC \citep{Akaike:1974, Sugiura:1978} and BIC \citep{Schwarz78} estimators are defined as

\begin{eqnarray}
{\rm AIC} = -2 \ln {\cal L}_{\rm max}+2k+\frac{2k(k+1)}{N-k-1} \label{eq:AIC}\;,\\
{\rm BIC} = -2 \ln {\cal L}_{\rm max}+k\ln{N}\;,
\end{eqnarray}
where ${\cal L}_{\rm max}$ is the highest likelihood function (proportional to minimum of $\chi^2$), $N$ is the number of observational data and $k$ is the number of free parameters. One can ignore the last term in R.H.S of Eq. (\ref{eq:AIC}) when the number of observational data $N$ is much more than the number of free parameters $k$. Among all models, the one that minimizes the AIC is considered to be the best one. If the difference between AIC of a given model and the best model is smaller than $4$ ($\Delta=AIC_{model}-AIC_{min}<4$), one concludes that the best fitted model and a given model are equally supported by the data. In the case of $4<\Delta <10$, observations still support the given model but less than the best one. Finally for $ \Delta>10$, observations basically do not support the given model comparing to the best model. The results of our analysis  are presented as follows:
\begin{itemize}
	\item  DF parameterization (case 1):  $\chi^{2}_{\rm min}=65.8$, $k=4$, $AIC=74.35$\;.\\
	\item  DF parameterization (case 2):  $\chi^{2}_{\rm min}=62.21$, $k=5$, $AIC=73.04$\;.\\
	\item $\Lambda$CDM model:  $\chi^{2}_{\rm min}=69.94$, $k=3$, $AIC=76.26$\;.
\end{itemize}
The above results show that the DF parameterization (case 2) has a lowest value of AIC. However since the difference between two cases of DF parameterization is {about}
$1.3$, we conclude both DF parameterizations are equally fitted to observational data in background level. Furthermore we see that  difference between the AIC value of $\Lambda$CDM cosmology and DF parameterization (case 2) is less
than $4$. Hence all models studied in this work are well fitted to observational data in background level. The best fit values of cosmological parameters are presented in Table (\ref{tab:res_all}). Also $1\sigma$ and $2\sigma$ confidence levels of cosmological parameters are shown in Fig. (\ref{figc2}).
\begin{table*}
	\centering
	\caption{Best fit values of cosmological parameters obtained in MCMC analysis using the geometrical data in background level.}
	\begin{tabular}{|c | c| c|c| } 
		\hline \hline
		Parameter& case 1 & case 2 &$\Lambda$CDM\\
		\hline
		$\Omega_{\rm b0}$ & $0.050^{+0.0008+0.002}_{-0.0008-0.002}$ & $0.051^{+0.0008+0.002}_{-0.001-0.002}$ &$0.049^{+0.0008+0.002}_{-0.0008-0.002}$\\
		\hline
		$ H_{0} $ & $68.9^{+0.64+1.3}_{-0.64-1.2}$& $68.36^{+0.65+1.3}_{-0.65-1.3}$&$69.2^{+0.69+1.4}_{-0.69-1.3}$\\
		\hline
		$ w_{0} $ &$-0.73^{+0.008+0.02}_{-0.008-0.02}$&$-0.69^{+0.02+0.04}_{-0.02-0.05}$ &-\\
		\hline
		$\alpha $ &$-51.3^{+13+22}_{-11-25}$&$-39.1^{+8.0+9.5}_{-8.4-11}$& 0\\
		\hline
		$w_{\rm de}$&$-1$&$-0.95^{+0.02+0.05}_{-0.03-0.05}$&$-1$\\
		\hline
		$\Omega_{\rm dm0} $ & 0.257&0.263&$0.246^{+0.008+0.02}_{-0.008-0.02}$ \\
		\hline
		$ \Omega_{de0} $ & 0.693&0.686 &0.705 \\
		\hline \hline
	\end{tabular}\label{tab:res_all}
\end{table*}

\subsection{Cosmological evolution}	

In this section based on the best fit values of cosmological parameters presented in Tab. (\ref{tab:res_all}), we depict the evolution of main cosmological quantities in the framework of DF parameterization. In Fig. (\ref{fig: 2}) we present the evolution of fractional energy densities for radiation, DM and DE. For all models, the Universe evolves from radiation dominated phase to matter dominated epoch and finally enters the late time DE dominated phase. The radiation-matter equality epoch for $\Lambda$CDM model occurs at $a_{eq} \approx 2.9\times10^{-4}$. For DF parameterization case 1 (case 2) it happens later at $a_{eq}\approx5.5\times10^{-4}$($a_{eq}\approx5.1\times10^{-4}$). In all models we see that DE starts to dominate the energy budget of Universe at $a\sim 0.75$. Notice that in the case of DF parameterization, due to interaction between DM and DE, the evolution of energy density of both DM and DE is different from the one in the $\Lambda$CDM scenario.

\begin{figure}
	\begin{center}
		\includegraphics[height=5cm,width=8cm]{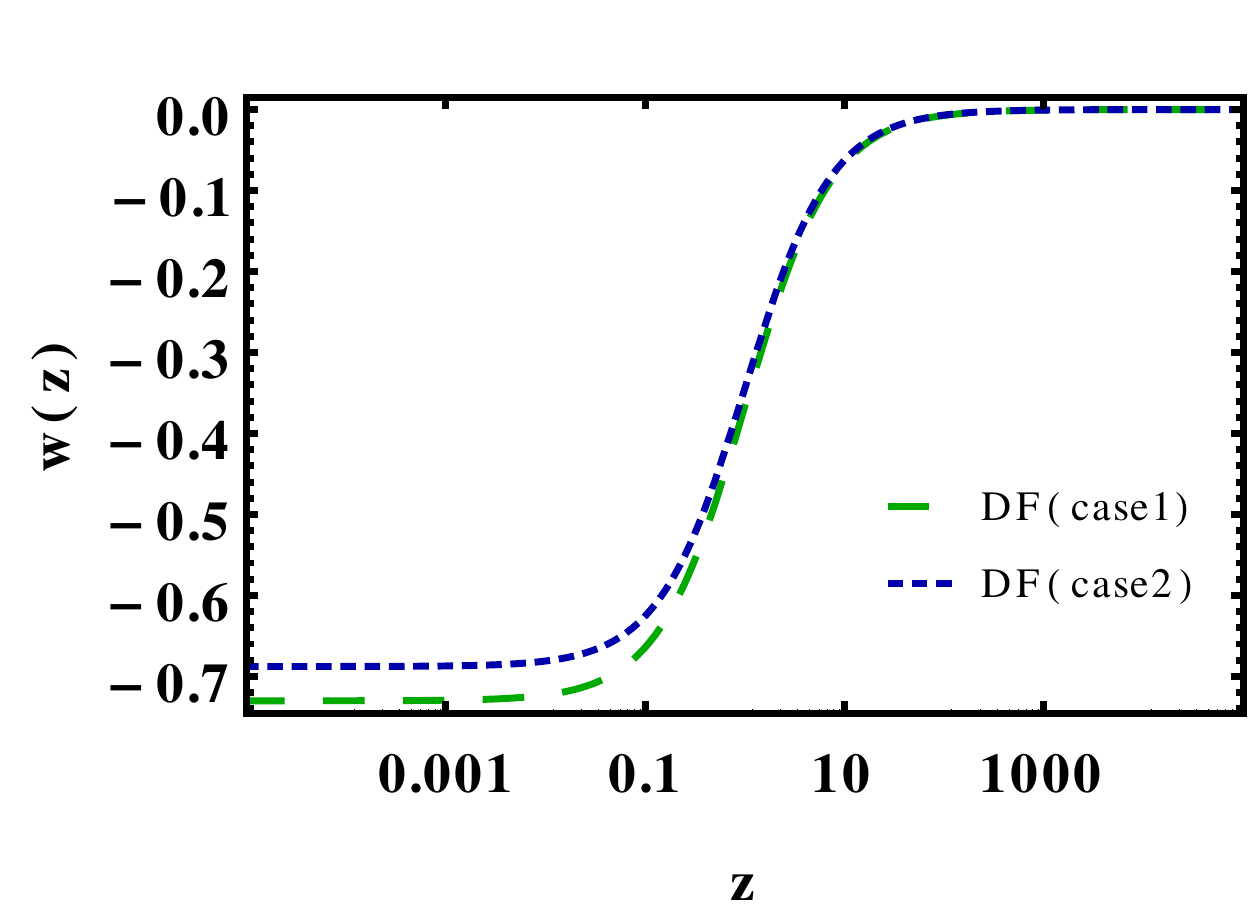}
		\includegraphics[height=5cm,width=8cm]{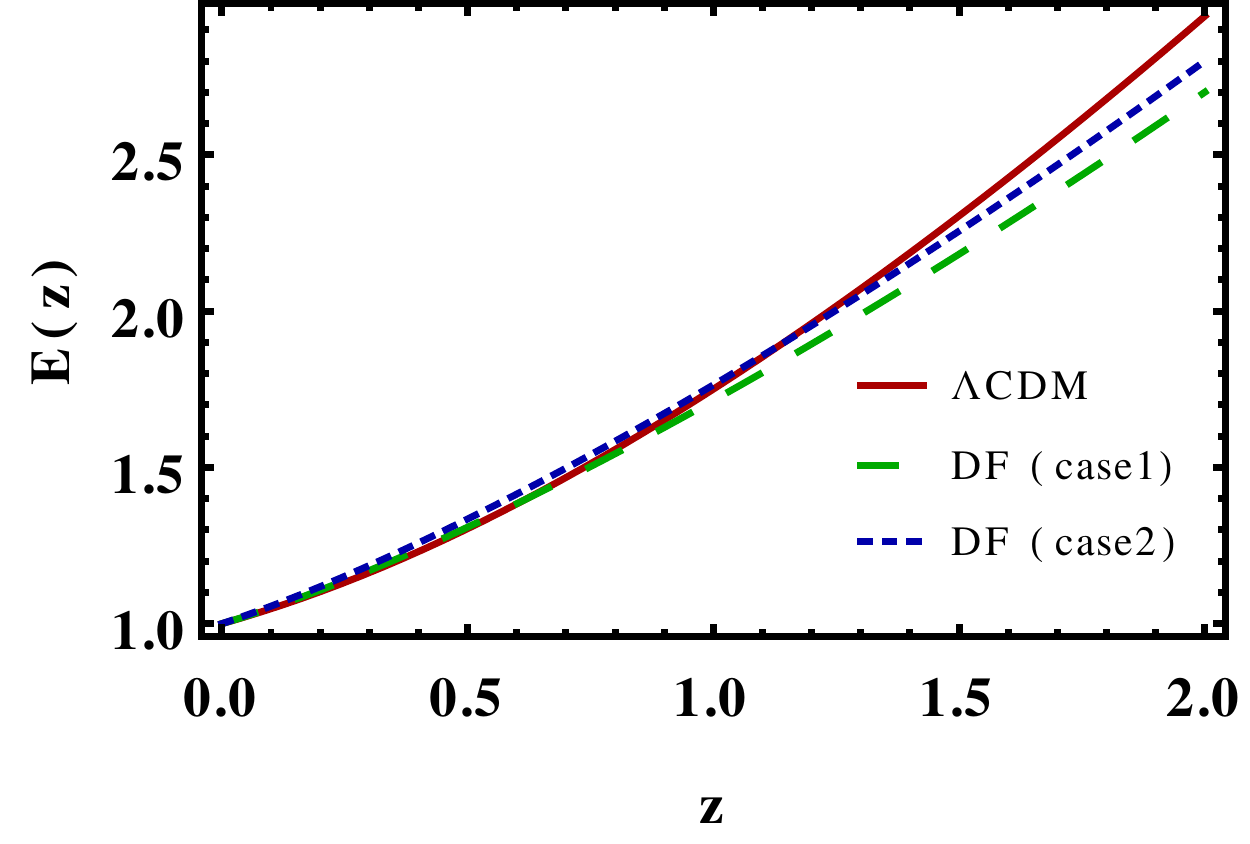}
		\includegraphics[height=5cm,width=8cm]{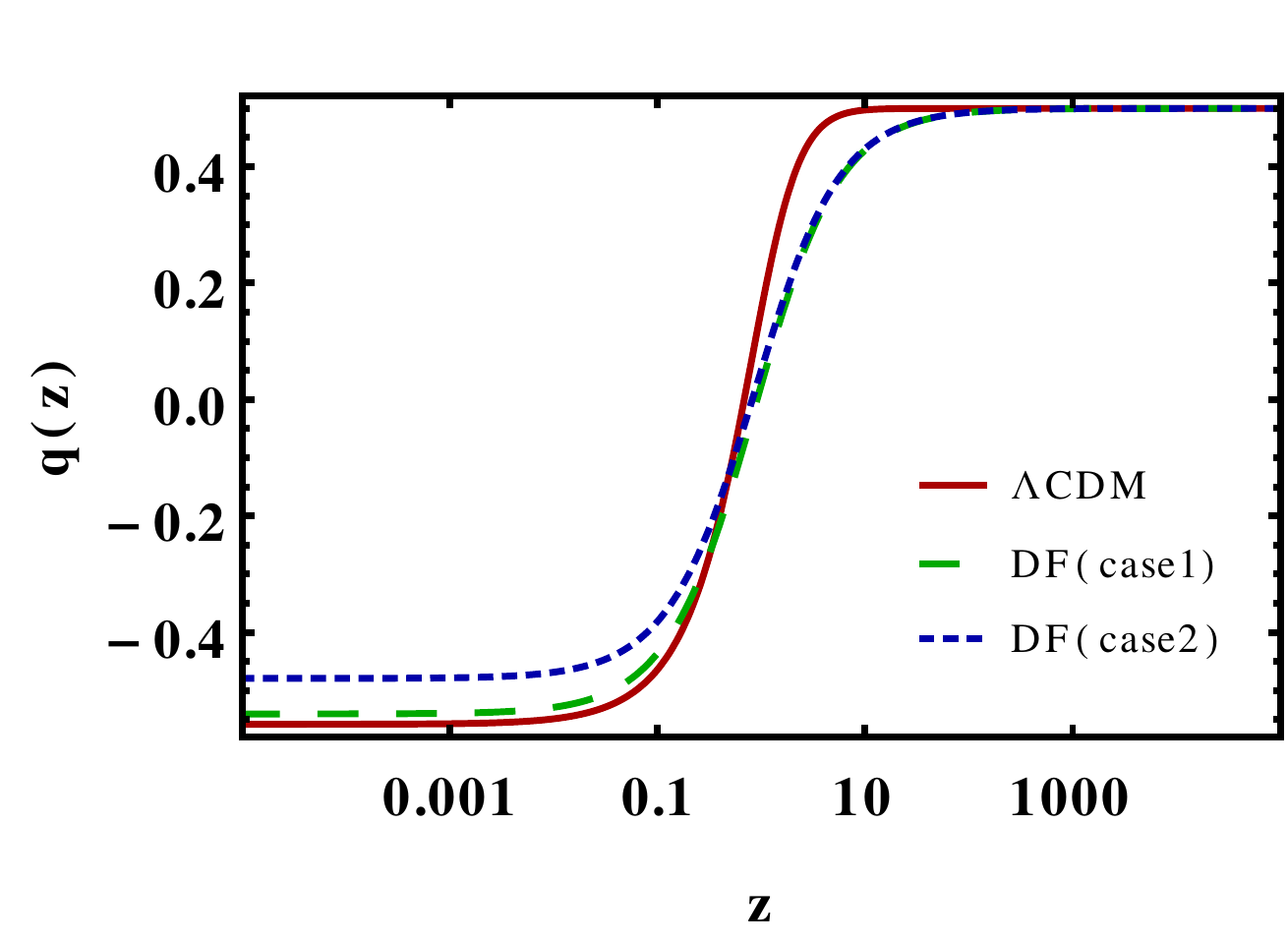}
		\caption{Top panel: Redshift evolution of EoS parameter, {$w$}, of DF parameterizations. Middle panel: redshift evolution of Hubble parameter $E$. Bottom panel: redshift evolution of deceleration parameter $q$. In all panels, case 1 and case 2 are shown {with} dashed and dotted curves respectively. The concordance $\Lambda$CDM cosmology is shown by red solid curve.}\label{fig:qz}
	\end{center}
\end{figure}

In Fig.\ref{fig:qz}, we show the evolution of the EoS parameter of DF $w$, the Hubble parameter $E$ and deceleration parameter $q$ as a function of redshift $z$. Notice that in case 2 we use the best fit values for cosmological parameters based on Tab. (\ref{tab:res_all}).  In top panel, for both cases, we observe that the EoS parameter of DF tends to zero at high redshifts representing the pressure less matter fluid $w_{\rm m}=0$ at high redshift Universe. While the redshift decreases, the EoS parameter decreases and reaches negative values smaller than $-1/3$, which is a necessary condition for DF to behave as DE at late times. In middle panel, the redshift evolution of Hubble parameter $E$ has been shown for both cases of DF parameterization. The $\Lambda$CDM case is shown for comparison. We see that at high redshifts the Hubble parameter for DF parameterizations is smaller than $\Lambda$CDM case. However at low redshifts, the Hubble parameter in DF parameterization (case 1) coincides the $\Lambda$CDM model. This result is expected since we assume the DE component of DF as cosmological constant $\Lambda$ with $w_{\rm de}=-1$. At low redshifts, when DE component of DF dominates its evolution, we expect that the Hubble parameter in DF (case 1) fits to the one from the $\Lambda$CDM cosmology. On the other hand, we observe that at low redshifts the Hubble parameter in case 2 is a little larger than $\Lambda$CDM cosmology. This is due to fact that we consider the DE part of DF as a quintessence DE with $w_{\rm de}>-1$. Finally in the bottom panel we present the evolution of a deceleration parameter $q(z) = -1 - \frac{\dot H}{H^2}$ for $\Lambda$CDM and DF parameterizations. The connection between $q(z)$ and $w_{de}$ is rather straightforward. From the Friedmann equations one finds

\begin{equation}\label{eq:qq}
\frac{\dot H}{H^{2}}=-\frac{3}{2}(1+w_{de}(z)\Omega_{de}(z))\;.
\end{equation}
which implies that
\begin{equation}\label{eq:q2}
q(z)=\frac{1}{2}+\frac{3}{2}w_{de}(z)\Omega_{de}(z)\;.
\end{equation}
In both $\Lambda$CDM model and DF parameterizations, $q$ tends to $\frac{1}{2}$ at early times as expected. By solving the $q(z_{tr})= 0$ we can obtain the transition redshift, namely the epoch at which the expansion of the Universe starts to accelerate. In particular, we find $z_{tr}= 0.9$ for DF parameterization (case 1), $z_{tr}= 0.87$ for DF parameterization (case 2) and  $z_{tr}= 0.66$ for $\Lambda$CDM. Hence $\Lambda$CDM model is entering the accelerating phase later than DF parameterizations. These results are in good agreement with the measured  $z_{tr}$ based on the cosmic  $H(z)$ data \cite{Farooq:2016zwm} \citep[see also][]{Capozziello:2014zda, Capozziello:2015rda}).\\
\section{GROWTH OF PERTURBATIONS  }\label{sec:5}
Beside the observational data in the background level it is important to study the features of different DE models using the observations in cluster scales. In fact most of the dynamical DE models introduced as parallel candidates of $\Lambda$ have the same behaviors at background cosmology. However they may have different evolution in cluster scales where we study the growth history of perturbations. Therefore the information from large scale structure formation is a powerful tool to examine different types of DE models. In this section we start from presenting the basic equations for the growth of matter perturbations in DF scenarios followed by the examination of the DF parameterizations against the observational growth rate data in perturbation level.
\subsection{basic equations}
We consider the scalar perturbations, which are developed at low redshifts in the era of structure formation. The perturbed line element in the conformal Newtonian gauge as
\begin{equation}
ds^2=a^2(\tau)[-(1+2\psi)d\tau^2+(1-2\phi)\delta_{ij}dx^idx^j ] \;,
\end{equation}
where $x$ is the spatial coordinates and $\psi$ and $\phi$ are the linear gravitational potentials. In the limit of GR and in the absence of anisotropic stresses one finds $\psi=\phi$. We use Latin letters $i, j,...$ for the spatial indexes $1, 2, 3$ and Greek letters $\mu, \nu,...$ for indexes $0, 1, 2$ and 3. In this formalism, metric is given by

\begin{equation}\label{eq:metric}
g_{\mu\nu} = a^2
\left({
	\begin{array}{cc} 
	-(1+2\psi)&0\\
	0&(1-2\phi) \, \delta_{ij}\\
	\end{array}}\right) \, .
\end{equation}
We can rewrite the perturbed metric in Eq. (\ref{eq:metric}) by separating $h_{\mu\nu}$ as $g_{\mu\nu}=\bar{g}_{\mu\nu}+h_{\mu\nu}$ where $\bar{g}_{00}=-a^2,\bar{g}_{ij}=a^2\delta_{ij}$ are the metric components describing the background and $h_{00}=-2a^2\phi$, $h_{ij}=-2a^2\phi \,\delta_{ij}$ are the perturbations of metric \citep{Marcondes:2016reb}. The energy-momentum tensor for a perfect fluid in an homogeneous and isotropic Universe reads 
\begin{equation}
\bar{T}_{\mu\nu}=\bar{P}\bar{g_{\mu\nu}}(\bar{\rho}+\bar{P})\bar{u}_{\mu}\bar{u}_{\nu}
\end{equation}
where, $u_{\mu}$ is the four-velocity. Let us decompose the energy-momentum tensor into $T_{\mu\nu}=\bar{T}_{\mu\nu}+\delta T_{\mu\nu}$, where $\bar{T}_{\mu\nu}$ and $\delta T_{\mu\nu}$ are background and perturbations respectively. The perturbed part of the energy momentum tensor is the following
\begin{equation}\label{eq:pert33}
\delta T_{\mu\nu}=(\delta\rho+\delta p)\bar{u}_{\mu}\bar{u}_{\nu}+(\bar{\rho}+\bar{P})(\delta u_{\mu}\bar{u}_{\nu}+\bar{u}_{\mu}\delta u_{\nu})-\delta p\delta_{\mu\nu}\;,
\end{equation}
{where different nonzero component of Eq. (\ref{eq:pert33})} are obtained as
\begin{eqnarray}
&&\delta T_0^0=-\delta\rho,\qquad \delta T_j^i=\delta p\delta_i^j,\qquad \delta T_{00}=a^2(\delta\rho+2\bar{\rho}\phi),\nonumber\\
&&\delta T_0^i=-a^{-1}(\bar{\rho}+\bar{P})\delta u^i,\qquad\quad \delta T_i^0=a^{-1}(\bar{\rho}+\bar{P})\delta u_i,\nonumber\\
&&\delta T_{0i}=\delta_{i0}=-a(\bar{\rho}+\bar{P})\delta_i \, , \ \  \delta T_{ij}=a^2(\delta p-2\bar{p}\phi)\delta_{ij}\;,
\end{eqnarray}
where the bars indicate that the quantities are unperturbed. Using the the perturbed metric $g_{\mu\nu}$ and the perturbed conservation equations, we can obtain the following evolution equation for the evolution of the matter perturbations \citep{Marcondes:2016reb, Bento:2004uh}

\begin{eqnarray}
&&-\dot{\delta}-\left[3\mathcal{H}(c_s^2-w_{de})-\frac{\bar{Q}_0}{\rho}\right]\delta-(1+w_{de})(\theta-3\dot{\phi})=\frac{\delta Q_0}{\rho}, \nonumber\\
&&\dot{\theta}+\left[\mathcal{H}(1-3w_{de})-\frac{\bar{Q_0}}{\rho}+\frac{\dot{w}_{de}}{1+w_{de}}\right]\theta-k^2\phi-\frac{c_s^2k^2}{1+w_{de}}\delta \nonumber\\
&&\qquad\qquad\qquad\qquad\qquad\qquad\qquad\qquad=\frac{ik^i\delta Q_i}{\bar{\rho}(1+w_{de})}\label{eqdel}
\end{eqnarray}
where overdot is a derivative with respect to conformal time, $\delta \equiv \delta \rho/\bar{\rho}$ is the density perturbation, $c_s^2\equiv\delta_p/\delta \rho$ is the sound speed of the DE, $k^i$ are the component of the wavevector in Fourier space, $\theta\equiv a^{-1}ik^j\delta u_j$ is the divergence of the velocity perturbation in Fourier space, $\delta Q_i$ are the perturbation of the exchange of energy-momentum in the perturbed conservation equations and $\mathcal{H}$ is the conformal Hubble parameter. Notice that $\bar{Q}_0$ in above equation is the exchange of energy between DM and DE at the background level and due to the homogeneity and isotropy of the Universe at the background level its spatial components are zero (see also Eq.\ref{eq:Q}).

\begin{figure*}
	\includegraphics[width=8cm]{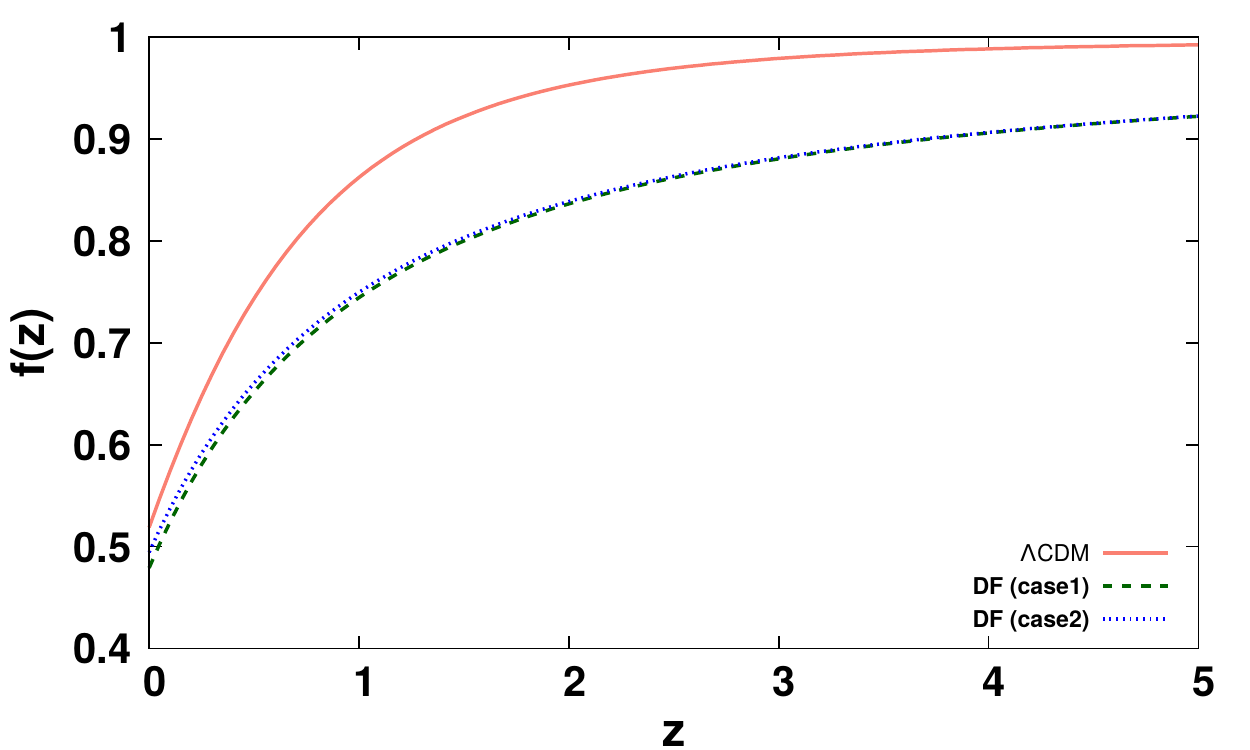}
	\includegraphics[width=8cm]{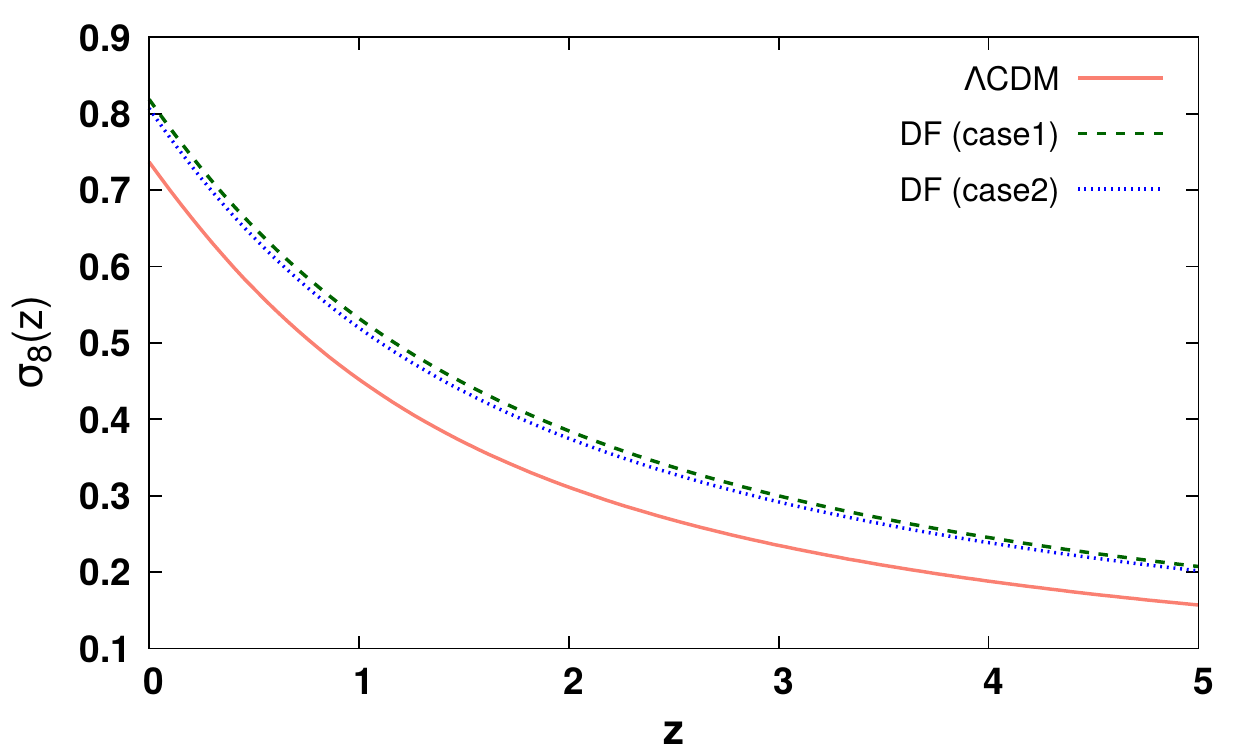}
	\caption{{Redshift evolution of matter growth rate function $f(z)$ (left panel) and  variance of perturbations $\sigma_8$ (right panel) in the context of DF cosmology. The background cosmological parameters are fixed using their values obtained in Tab. (\ref{tab:res_all}). The interaction parameter $\xi$ and $\sigma_8(z=0)$ are fixed using their constrained values} in Tab. (\ref{tab:all}). The concordance $\Lambda$CDM cosmology is shown by solid curve.}
	\label{fig:growth}
\end{figure*}
\begin{table*}
	\centering
	\caption{ Numerical results for different DF parameterizations and $\Lambda$CDM model obtained from the statistical MCMC analysis using the cosmological growth rate data in cluster scales. The best fit values of cosmological parameters $\xi$ and $\sigma_8(z=0)$ with their $1\sigma$ and $2\sigma$ confidence levels are shown in tow first column. The minimum of least square function $\chi^2$ and minimum of AIC value are shown in two last rows.}
	\begin{tabular}{|c | c|  c|c|}
		\hline \hline
		Parameter& case 1 & case 2 & $\Lambda$CDM\\
		\hline
		$\xi(10^{-4})$ & $7.2^{+0.003+0.006}_{-0.003-0.006}$ &$7.3^{+0.17+0.3}_{-0.17-0.3}$&0\\
		\hline
		$\sigma_{\rm 8}$ & $0.802^{+0.016+0.02}_{-0.015-0.02}$ &$0.81^{+0.02+0.03}_{-0.02-0.04}$&$0.74^{+0.02+0.03}_{-0.02-0.04}$\\
		\hline
		$\chi^2_{min}$ & $7.87$ &7.68&8.11\\
		\hline
		$AIC_{min}$ & 12.51 & 12.31 & 10.31\\
		\hline \hline
	\end{tabular}\label{tab:all}
\end{table*}

\begin{figure*}
	\includegraphics[width=8cm]{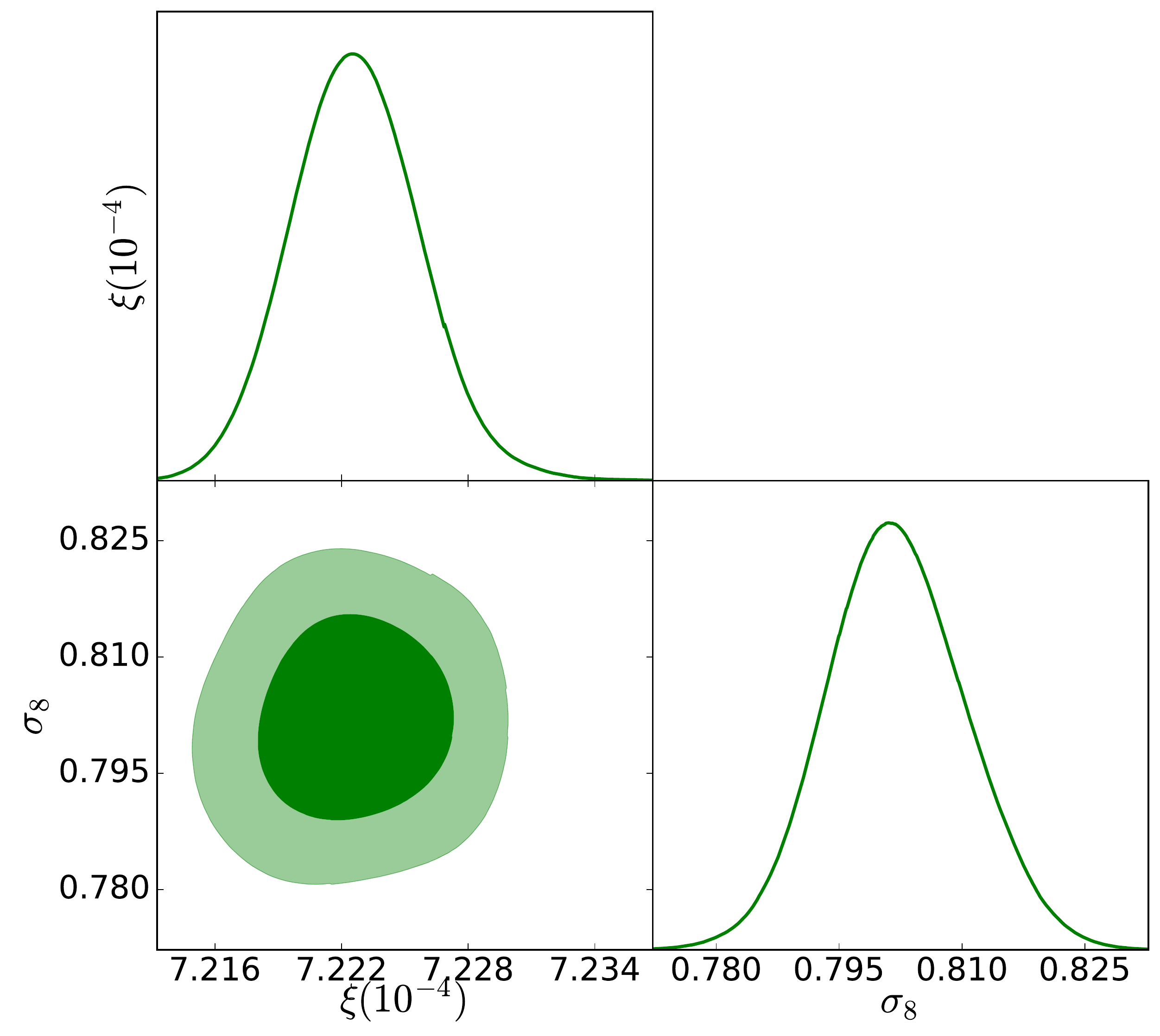}
	\includegraphics[width=8cm]{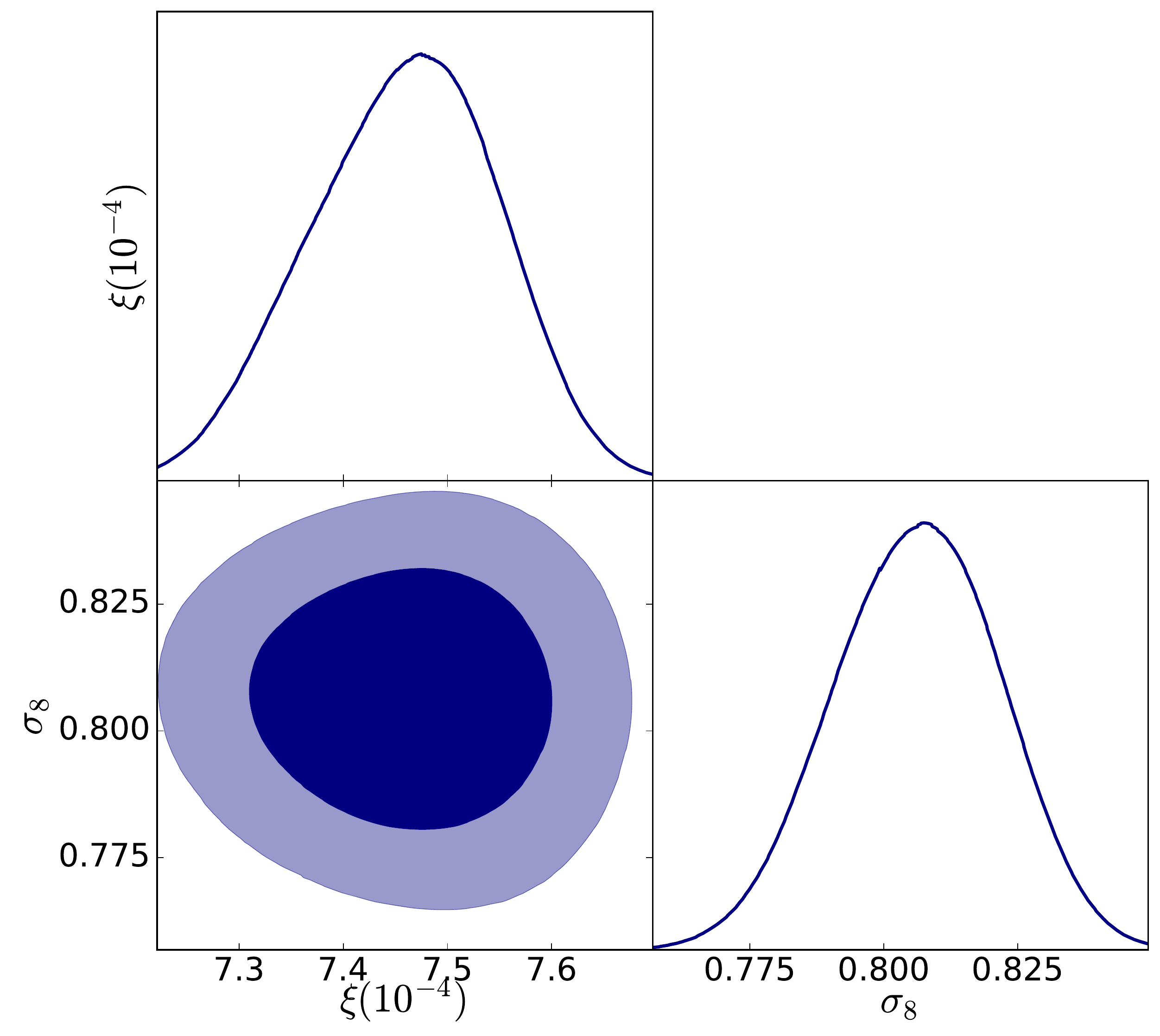}
	\caption{The $1\sigma$ and $2\sigma$ confidence contours and maximum likelihood functions in $\xi$-$\sigma_8(z=0)$ plane for case 1 (left panel) and case 2 (right panel) of DF parameterization.}
	\label{figc3}
\end{figure*}
\begin{figure*}
	\includegraphics[width=8cm]{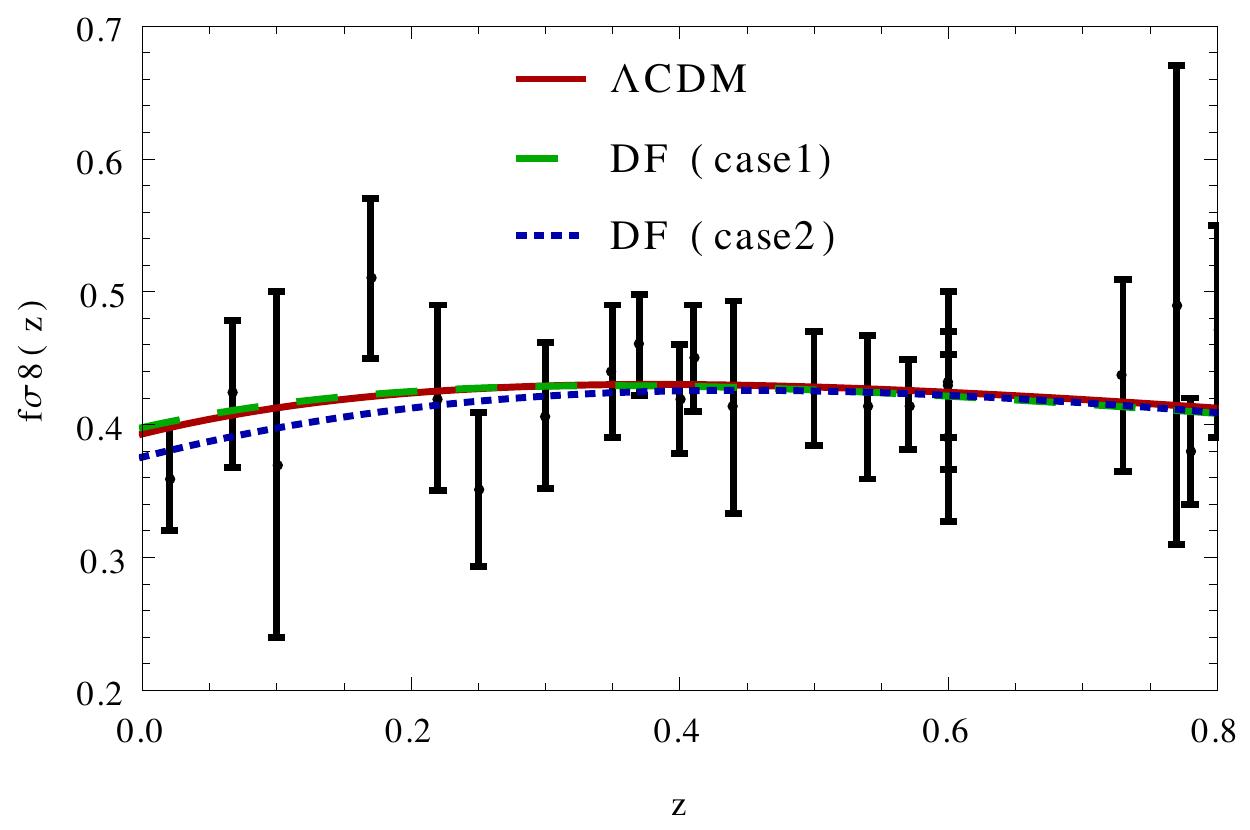}
	\caption{Theoretical predicted $f(z)\sigma_8(z)$ using the best fit values of cosmological parameters in Tabs. (\ref{tab:res_all} \& \ref{tab:all}) compared to observational growth rate data points. DF parameterizations case 1 and case 2 are shown by dashed and dotted curves respectively. The concordance $\Lambda$CDM cosmology is shown by red solid curve.}
	\label{figfs8}
\end{figure*}
Using the perturbed Poisson equation in the Fourier space, one finds
\begin{equation}
\left(1+\frac{3\mathcal{H}^2}{k^2}\right)k^2\phi=-3\mathcal{H}\dot{\phi}-4\pi Ga^2(\rho_{m}\delta_{m}+\rho_{de}\delta_{de}),\label{eqpoi}
\end{equation}
where $\delta_m$ is the density perturbation of pressure-less matter (baryons+dark matter) and $\delta_{de}$ is the DE density perturbation. We focus on the growth of perturbations with wavelength much smaller than the horizon ($k \gg \mathcal{H}$). In this limit we can use the Pesudo-Newtonian cosmology and neglect the time variation of gravitational potential. Hence we can ignore the second term of the left-hand side of Eq. (\ref{eqpoi}) and the term proportional to $\dot{\phi}$ at the right hand side. Also due to large sound horizon of DE, the DE perturbations $(\delta_{de}=0)$ are expected to be negligible on sub-horizon scales \citep{Duniya:2013eta}. Therefore, the Poisson equation reduces to
\begin{equation}
k^2\phi=-4\pi Ga^2\rho_{M}\delta_{m}=-\frac{3}{2}\mathcal{H}^2\Omega_{m}\delta_{m},\label{eqpoif}
\end{equation}
 where $\Omega_m=\Omega_{dm}+\Omega_b$ is the sum of fractional density parameter of dark matter and baryons.
Combining Eq. (\ref{eqdel}) with the Poisson equation (\ref{eqpoif}), we get
\begin{eqnarray}\label{eq:perturbed22}
&&\dot{\delta}_{m}+3\mathcal{H}\xi\frac{\Omega_{de}}{\Omega_{m}}{\delta}_{m}+\theta_{m}=0\nonumber\\
&&\dot{\theta}_{m}+\mathcal{H}(1+3\xi\frac{\Omega_{de}}{\Omega_{m}})\theta_{m}+\frac{3}{2}\mathcal{H}^2\Omega_{m}\delta_{m}=0 \, . \label{eqevo}
\end{eqnarray}
{Notice that here we use the reduced form of phenomenological equation (\ref{eq:Q}) as $Q=\bar{Q}_0=-3\xi \mathcal{H}\rho_{de}=-3\xi \mathcal{H}\frac{\Omega_{de}}{\Omega_{m}}\rho_{m}$, by setting $\xi_1=0$. Since we are neglecting DE clustering, we also ignore perturbations of $Q$ to derive Eqs.\ref{eq:perturbed22} \citep[see also][]{Marcondes:2016reb}}. By eliminating $\theta_{m}$ from the system of Eq. (\ref{eqevo}) and changing the variables from conformal time to physical time according to  $\mathcal{H}=aH$, $\frac{d}{d\tau}=a\frac{d}{dt}$ and $\frac{d^2}{d\tau^2}=a^2(\frac{d^2}{dt^2}+H\frac{d}{dt})$, we obtain the following equation
\begin{eqnarray}
\frac{d^2\delta_{m}}{dt^2}+2\left(H+3\xi\frac{\Omega_{de}}{\Omega_{m}}\right)\dot{\delta}_{m}-\frac{3}{2}H^2\Bigg[\Omega_{m}-2\xi\frac{\Omega_{de}}{\Omega_{m}}\times \nonumber\\
\left(1+\frac{\dot{H}}{aH^2}+3\xi\frac{\Omega_{de}}{\Omega_{m}}-\frac{\dot{\Omega}_{de}}{H\Omega_{m}\Omega_{de}}\right)\Bigg]\delta_{m}=0
\end{eqnarray}
Changing the time derivative into derivative with respect to scale factor $a$ ($\frac{d}{dt}=aH\frac{d}{da}$), we get
\begin{equation}
\delta_{m}^{''}+A_{m}\delta_{m}^{'}+B_{m}\delta_{m}=S_m\;,\label{ddm1}
\end{equation}
where coefficients $A_{m}$, $B_{m}$ and $S_m$ are written as 
\begin{eqnarray}
&& A_{m}=\frac{3}{a}+\frac{H'}{H}+\frac{6\xi}{a}\frac{\Omega_{de}}{\Omega_{m}};\nonumber\\
&& B_{m}=-\frac{3}{2a^2}\left[-2\xi\frac{\Omega_{de}}{\Omega_{m}}\left(1+\frac{H'}{H}+3\xi\frac{\Omega_{de}}{\Omega_{m}}-\frac{\Omega'_{m}}{\Omega_{m}\Omega_{de}}\right)\right]\nonumber\\
&&S_m=-\frac{3}{2a^2}\Omega_{m}\delta_{m}
\end{eqnarray}
Notice that by putting $\xi=0$ the standard equation for the evolution of matter perturbations can be recovered. Now we numerically solve Eq. (\ref{ddm1}) to obtain the evolution of the growth of matter perturbations in DF cosmology. Concerning the initial conditions, we set the initial scale factor $a_{\rm i}=0.0005$ ($z_{\rm i}=2000$) which means that we are deep enough in the early matter dominated era. We use $\delta_{\rm mi}=8 \times 10^{-5}$, which guarantees that the linear regime ($\delta_m<1$) of perturbations at the present time. The background cosmological parameters have been used from the best fit values presented in Tab. (\ref{tab:res_all}). In addition we set the interaction parameter $\xi$ to constrained value $\xi=7\times 10^{-4}$ obtained in Tab. (\ref{tab:all}). Once the matter perturbation $\delta_{m}(z)$ is obtained, we calculate the evolution of the growth rate function $f=\frac{d\ln\delta_{\rm m}}{d\ln a}$ and the mass variance of matter perturbations $\sigma_8$ within the sphere of $R_8= 8h^{-1}Mpc$.  The variance of perturbations within $R_8$ at redshift $z$ reads $\sigma_8(z)=D(z)\sigma_8(z=0)$, where $D(z)=\delta_{m}(z)/\delta_{m}(z=0)$ is the linear growth factor of matter perturbations and $\sigma_8(z=0)$ is the present value of variances. For models discussed in this work we fix $\sigma_8(z=0)$ from the constrained values in Tab. (\ref{tab:all}).

In left panel of Fig. (\ref{fig:growth}), we show the evolution of growth rate function $f$ as a function of cosmic redshift $z$. One can see that DE decreases the amplitude of matter perturbations at low redshift. We observe that in both cases of DF models and concordance $\Lambda$CDM cosmology, the growth rate of matter perturbations is suppressed due to the effect of DE component at low redshifts. Notice that at high redshifts the influence of DE on the growth of perturbations is negligible and consequently the growth function goes to unity, which corresponds to the matter dominated Universe. We conclude that the suppression of the amplitude of matter fluctuations in DF cosmologies starts sooner comparing to the standard $\Lambda$CDM model. This result can be extracted from Fig.(\ref{fig: 2}) in which the fractional energy density of DE at higher redshifts calculated in $\Lambda$CDM model vanishes sooner than in the case of DF cosmology. Hence the non-vanishing DE at higher redshifts suppresses the growth of matter fluctuations at earlier times. 

In right panel of Fig. (\ref{fig:growth}), we show the evolution of $\sigma_8(z)$ as a function of redshift $z$ computed for DF and $\Lambda$CDM models. Note, that the variance of perturbations in both DF cosmologies and $\Lambda$CDM model grows with a scale factor. Moreover, opposite to the behavior of growth rate function, the variance of perturbations in DF models is larger than the one in the case of  $\Lambda$CDM universe.

\subsection{growth rate data} 
\label{sect:growthdata}    

In this section we calculate the theoretical value of $f(z)\sigma_8(z)$ in the context of DF cosmology. Using the observational growth rate data \citep{Nesseris:2017vor}, we perform a statistical least square analysis to compute 
\begin{equation}\label{eq:chigr}
\chi^2_{\rm gr}=\Sigma_{i=1}^{N} {\frac{[f\sigma_8^{(th)}(z_{i})-{f}\sigma_{8}^{(obs)}(z_{i})]}{\sigma_i^2}} \, ,
\end{equation}
where $\sigma_i$ are corresponding uncertainties, ``$obs$'' stands for the observed data and ``$th$'' denotes the theoretical prediction in DF cosmology. The growth rate data used in this analysis comes from 18 distinct data points for $f\sigma_8(z)$ \citep{Nesseris:2017vor}. Here we consider the interaction parameter $\xi$ and mass variance $\sigma_8(z=0)$ as free parameters, which can be constrained by growth rate data. We fix the other cosmological parameters using the best fit values presented in Tab. (\ref{tab:res_all}). In MCMC analysis (performed using the growth rate data) the statistical vector $p$ contains two free parameters ($\xi$,$\sigma_8(z=0)$) for DF cosmology (both case 1 \& 2) and $\sigma_8(z=0)$ for concordance $\Lambda$CDM cosmology. Our results in this analysis are presented in Tab. (\ref{tab:all}). We show that the $\Lambda$CDM model, which appears to have the lowest AIC value, is the best model in cluster scales. However since the difference between AIC of DF parameterizations (both case 1 \& 2) and $\Lambda$CDM model is lower than $3$, we conclude that both cases of DF parameterizations are fitted to growth rate data as well as in the case of the $\Lambda$CDM cosmology. In Fig. (\ref{figc3}) we visualize the $1\sigma$ and $2\sigma$ confidence levels in $\xi-\sigma_8$ plane for DF parameterization case 1 (left panel) and for case 2 ( right panel). In Fig. (\ref{figfs8}), we show the theoretically predicted $f(z)\sigma_8(z)$ for DF parameterizations using the best fit cosmological parameters presented in Tabs. (\ref{tab:res_all} \& \ref{tab:all}). We see that both DF parameterizations are well fitted with observational growth rate data  and that the fit is as good as in the case of $\Lambda$CDM cosmology. This result is comparable with the implications of Fig. (\ref{fig:growth}), in which the predicted growth rate function $f(z)$ in DF cosmologies is lower than that in $\Lambda$CDM model (see left panel), while the quantity $\sigma_8(z)$ calculated in DF models is higher than the same quantity in $\Lambda$CDM universe (right panel). Hence, one can conclude that the production of growth rate function and variance of perturbation, i.e., $f(z)\sigma_8(z)$, of DF cosmology is compatible with that of the one in $\Lambda$CDM model (see Fig. (\ref{figfs8})).

\section{Conclusion}\label{sec:6}

In this paper we have proposed a novel parameterization of a dark fluid, which may include both DM and DE. In the Sec. \ref{sec: 2}, we have considered the barotropic equation of state parameter $w(a)$ as a general polynomial function with a scale factor as a variable. We have shown that the existence of a stationary point of $w(a)$ secures the lack of divergences of $w$ in the far future. For a simple example of such a barotropic parameter with a stationary point, we have shown that it can describe the smooth transition between initial zero barotropic parameter $w_i=0$ and negative DE equation of state parameter $w_{\rm de}<-1/3$, therefore it may be used to unify DM and DE in one dark fluid.

In the Sec. \ref{sec:3} we have investigated the DF parameterization in two cases: for constant and dynamical energy density of the DE. We have included the constraints on the background evolution of the Universe using data from JLA supernovae, BAO, CMB, BBN and Hubble expansion. We have implemented the MCMC statistical analysis for two considered DE scenarios and we compared the results with the $\Lambda$CDM model. We have proven that both DF parameterizations are well fitted to observations as equally as concordance $\Lambda$CDM model. For all considered models we have also founded the redshift, for which the Universe starts to accelerate. The results vary from $z_{tr} = 0.9$ (for the case 1) to $z_{tr} = 0.66$ for the $\Lambda$CDM, all in good agreement with recent works \citep{Capozziello:2014zda,Capozziello:2015rda,Farooq:2016zwm,Rezaei:2017yyj}.

In the sec. \ref{sec:5}, we have investigated the growth rate of matter perturbations in the context of unified DF cosmology. We have shown that in this model the DE component, like $\Lambda$ sector in the standard $\Lambda$CDM model, can suppress the amplitude of matter perturbations at low redshift while its effects on the growth rate are negligible at high redshift epochs. We have also proven that both cases of DF parameterization are consistent with growth rate data in cluster scales as equally as the concordance $\Lambda$CDM model.

\section{Acknowledgments}

MA was supported by the Iuventus Plus grant No. 0290/IP3/2016/74 from the Polish Ministry of Science and Higher Education. 

 \bibliographystyle{apsrev4-1}
  \bibliography{ref}

\end{document}